\documentclass[fleqn,usenatbib]{mnras}

\usepackage{mathptmx}

\usepackage[T1]{fontenc}
\usepackage{ae,aecompl}

\usepackage{graphicx}	
\usepackage{amsmath}	
\usepackage{amssymb}	
\usepackage{float}
\usepackage{hyperref}
\usepackage{subcaption}
\usepackage{multicol}
\usepackage{array}
\usepackage{threeparttable}
\usepackage[usenames, dvipsnames]{color}
\renewcommand{\deg}{^\circ}

\newcolumntype{L}{>{\centering\arraybackslash}m{3.3cm} }

\title[]{Search for very-high-energy gamma-ray emission from the microquasar Cygnus\,X-1 with the MAGIC telescopes}

\author[M.~L.~Ahnen~et.~al.]{
M.~L.~Ahnen$^{1}$,
S.~Ansoldi$^{2,25}$,
L.~A.~Antonelli$^{3}$,
C.~Arcaro$^{4}$,
A.~Babi\'c$^{5}$,
B.~Banerjee$^{6}$,\newauthor
P.~Bangale$^{7}$,
U.~Barres de Almeida$^{7,26}$,
J.~A.~Barrio$^{8}$
% ...
(Affiliations can be found after the references)
}

\date{Accepted XXX. Received YYY; in original form ZZZ}

\pubyear{2017}

\begin{document}
\label{firstpage}
\pagerange{\pageref{firstpage}--\pageref{lastpage}}
\maketitle

\begin{abstract}
The microquasar Cygnus\,X-1 displays the two typical soft and hard X-ray states of a black-hole transient. During the latter, Cygnus X-1 shows a one-sided relativistic radio-jet. Recent detection of the system in the high energy (HE; $E\gtrsim60$ MeV) gamma-ray range with \textit{Fermi}-LAT associates this emission with the outflow. Former MAGIC observations revealed a hint of flaring activity in the very high-energy (VHE; $E\gtrsim100$  GeV) regime during this X-ray state. We analyze $\sim97$ hr of Cygnus X-1 data taken with the MAGIC telescopes between July 2007 and October 2014. To shed light on the correlation between hard X-ray and VHE gamma rays as previously suggested, we study each main X-ray state separately. We perform an orbital phase-folded analysis to look for variability in the VHE band. Additionally, to place this variability behavior in a multiwavelength context, we compare our results with \textit{Fermi}-LAT, \textit{AGILE}, \textit{Swift}-BAT, \textit{MAXI}, \textit{RXTE}-ASM, AMI and RATAN-600 data. We do not detect Cygnus X-1 in the VHE regime. We establish upper limits for each X-ray state, assuming a power-law distribution with photon index $\Gamma=3.2$. For steady emission in the hard and soft X-ray states, we set integral upper limits at 95\% confidence level for energies above 200 GeV at $2.6\times10^{-12}$~photons cm$^{-2}$s$^{-1}$ and $1.0\times10^{-11}$~photons cm$^{-2}$s$^{-1}$, respectively. We rule out steady VHE gamma-ray emission above this energy range, at the level of the MAGIC sensitivity, originating in the interaction between the relativistic jet and the surrounding medium, while the emission above this flux level produced inside the binary still remains a valid possibility. 
\end{abstract}



\section{Introduction}
\label{section1}
Cygnus\,X-1 is one of the brightest and best studied X-ray sources in our Galaxy and the first identified stellar-mass black hole (BH) X-ray binary system. Discovered in early stage of the X-ray astronomy \citep{1972Natur.235..271B}, the system is located in the Cygnus region ($l=71.32\deg$, $b=+3.09\deg$) at a distance of $1.86^{+0.12}_{-0.11}$ kpc from the Earth \citep{Reid2011}. It is comprised of a ($14.81\pm0.98$) M$_{\odot}$ BH and a O9.7 Iab type supergiant companion star with a mass of ($19.16\pm1.90$) M$_{\odot}$ \citep{Orosz2011}. Nevertheless, the most plausible mass range of the donor star has been recently increased to 25-35 M$_{\odot}$ by \cite{2014MNRAS.440L..61Z}. The orbit is almost circular ($e=0.18$, \citealt{Orosz2011}) with a  $\sim 5.6$ d period ($5.599829\pm0.000016$, \citealt{Brocksopp1999}) and an inclination angle of the orbital plane to our line of sight of $(27.1\pm0.8)\deg$ \citep{Orosz2011}. The superior conjunction of the compact object, when the companion star is interposed between the BH and the observer, corresponds to orbital phase 0, assuming the ephemerides $T_{0}$=52872.788 HJD taken from \cite{2008ApJ...678.1237G}. 
The assumption that Cyg\,X-1 ranks among the microquasars was accepted after the detection, by the VLBA instrument, of a highly collimated one-sided relativistic radio-jet that extends $\sim15$ mas from the source (opening angle $<2\deg$ and velocity $\geq$0.6c, \citealt{Stirling:2001xb}). This jet is thought to create a 5 pc diameter ring-like structure observed in radio that extends up to $10^{19}$ cm from the BH \citep{2005Natur.436..819G}. 

The compact object accretes material through an accretion disk from the supergiant companion star. Cyg\,X-1 displays the two principal spectral X-ray states of a BH transient system that can be divided according to the dominance level of a power-law component and a thermal component at lower keV energies \citep{Tanaka:1996ti}: the hard state (HS) and the soft state (SS; \citealt{Esin1998}). The HS is dominated by a power-law photon distribution (with $\Gamma\sim1.4-1.9$) with a high-energy exponential cutoff at $\sim150$ keV \citep{Gierlinski1997}. It is thought to be produced by Comptonization of thermal photons from the accretion disk by high-energy electrons in the so-called corona, hot ($T\sim10^{9}$ K) plasma at the inner region of the accretion flow \citep{Coppi1999}. The thermal component is negligible during this state. On the other side, the spectral energy distribution of the SS is characterized by a dominant thermal component that peaks at $kT\sim1$ keV, emitted mainly in the inner region of the accretion disk that extends down to the last stable orbit, and a softer power-law tail. In the transition between these two principal X-ray spectral states, an intermediate state (IS) occurs that lasts only a few days \citep{Grinberg2013}. For a comprehensive review on the subject, see \cite{Done2007}.

The X-ray emission from Cyg\,X-1 during its spectrally hard state is correlated with the radio emission originating in the relativistic jets \citep{2003MNRAS.344...60G}. During the HS, the jet is persistent and steady, except for some unusual flares \citep{Fender:2006kw} whereas, once the source enters in the SS, the jet may become unstable giving rise to a rapid jet Lorentz factor increase that originates an internal shock inside the outflow before being disrupted \citep{Fender2004}. In this SS state, the radio emission is not detected \citep{Brocksopp:1999xs}. 

Generally, X-ray binaries experience flux periodicity with the orbital period at different wavelengths. Cyg\,X-1 shows this kind of modulation both in X-ray and radio wavelengths (\citealt{Wen1999}, \citealt{Brocksopp:1999xs}, \mbox{\citealt{SzostekZdziarski2007}}), originating in absorption and/or scattering of the radiation from the compact object by the wind of the donor star. Besides this orbital modulation, several X-ray binary systems also present flux variations at much longer periods than their respective orbital period, known as superorbital modulation, that is thought to be caused by the precession of the accretion disk and/or jet \citep{Poutanen2008}. Cyg\,X-1 shows an X-ray superorbital period of $\sim 300$ d, as suggested by \cite{Rico2008} and confirmed by \cite{ZdziarskiPooley2011}. 

Observations with COMPTEL during Cyg\,X-1 SS suggested, for the first time, the existence of a non-thermal spectral component beyond MeV \citep{McConnell2002}. This result gave rise to an increase of the interest for this source in the gamma-ray regime. Nevertheless, observations with \textit{INTEGRAL} could not confirm the existence of this MeV tail in the SS, but probed, in turn, the presence of non-thermal hard emission during the HS \citep{2015ApJ...807...17R}. \textit{INTEGRAL}/IBIS also reported a hard tail in the HS which was shown to be polarized in the energy range of 0.4-2 MeV at a level of $\sim70$\% with a polarization angle of ($40.0\pm14.3$)$\deg$ (\citealt{2011xau5.confE..34L}, \citealt{2012ApJ...761...27J}). 

The recent detection of high energy (HE; $E\geqslant60$ MeV) gamma rays from Cyg\,X-1 associated with the jets \citep{Zanin2016}, using 7.5 yr of PASS8 \textit{Fermi}-LAT data, provided the first significant detection of HE gamma rays in a BH binary system. This steady emission was previously hinted by \cite{2013MNRAS.434.2380M}. \cite{Zanin2016} show that Cyg\,X-1 displays persistent HE emission during the HS (at $7\sigma$). This emission was suggested to be produced outside the corona (at distances $>2\times10^{9}$~cm from the BH), most likely from the jets. This was also pointed out by the fact that the detection happens only in the HS. A hint of gamma-ray orbital modulation was also found: the HE emission seems to happen when Cyg\,X-1 was at phases that cover the superior conjunction (between 0.75 and 0.25). This modulation, if confirmed, excludes the interaction  between the jets and the surrounding medium at large scales as the GeV emitter and suggests anisotropic inverse Compton (IC) on stellar photons, which constrains the emission region to $10^{11}$--$10^{13}$~cm from the compact object. The overall spectrum from \cite{Zanin2016} is well fitted by a power-law function with a photon index of $\Gamma=2.3\pm0.1$ and extends from 60 MeV up to $\sim20$ GeV. Besides this persistent emission, the source underwent 3 preceding episodes of transient emission detected by \textit{AGILE}. The first two flaring events occurred during the HS on October 16 2009, with an integral flux of ($2.32\pm0.66)\times10^{-6}$ ~photons cm$^{-2}$ s$^{-1}$ between 0.1 and 3 GeV \citep{2010ApJ...712L..10S}, and on March 24 2010, with an integral flux of $2.50\times10^{-6}$~photons cm$^{-2}$ s$^{-1}$ for energies above 100 MeV \citep{2010ATel.2512....1B}. The third one, on June 30 2010 with a flux of ($1.45\pm0.78)\times10^{-6}$~photons cm$^{-2}$ s$^{-1}$ also for energies above 100 MeV \citep{AGILE2013}, took place during the IS when the source was leaving the HS but just before an atypical radio flare \citep{2012MNRAS.419.3194R}. Each of these episodes lasted only 1--2 days. 

Although the gamma-ray spectrum does not seem to harden above $\sim20$ GeV, former MAGIC observations in the very high energy (VHE; $E\geqslant100$  GeV) band yielded a 4.1$\sigma$ evidence for VHE activity from the Cyg\,X-1 direction (referred as \textit{MAGIC hint}, hereafter). These MAGIC observations were carried out between June and November 2006 for 40 hr with the first stand-alone MAGIC telescope (MAGIC\,I). Although no significant excess for steady gamma-ray emission was found, during the daily analysis the \textit{MAGIC hint} was detected after 80 min on September 24 2006 (MJD= 54002.96; \citealt{Albert2007}), at the maximum of the superorbital modulation of the source and simultaneously with the rising phase of a hard X-ray flare detected by \textit{INTEGRAL}, \textit{Swift}/BAT and \textit{RXTE}-ASM \citep{Malzac2008}. The energy spectrum computed for this day is well reproduced by a power-law of d$\phi$/d$E$=($2.3\pm0.6)\times10^{-12}(E/1$TeV$)^{-3.2\pm0.6}$TeV$^{-1}$cm$^{-2}$s$^{-1}$. The VERITAS Collaboration observed Cyg\,X-1 in 2007 without any significant detection \citep{VERITASICRC09}.

In this paper we report observations of Cyg\,X-1 performed with the MAGIC telescopes between 2007 and 2014. Cyg\,X-1 was observed focusing on the HS concurrently with a high hard X-ray flux in order to perform observations under the same conditions as those during the \textit{MAGIC hint}. Section 2 describes the technical conditions of the MAGIC telescopes for each period, the observations of the source and data analysis. Section 3 reports the results obtained with MAGIC. We searched for steady gamma-ray emission using the entire data sample as well as splitting the data according to the spectral state. We also looked for signal in an orbital phase-folded analysis in both main X-ray states. Due to the variability that Cyg\,X-1 presents, daily analysis was also carried out and studied within a multiwavelength context. The physical interpretation and conclusions are given in Section 4. 

\begin{table*}    
\begin{center}
\caption{\textit{From left to right:} date of the beginning of the observations in calendar and in MJD, effective time after quality cuts, zenith angle range, X-ray spectral state and observational conditions (see Section \ref{section2}). Horizontal lines separate different observational modes along the campaign. During MJD 54656, 54657 and 54658, data under different observational modes were taken.}         
\label{table:1}      

	\begin{tabular}{ c c c c c c}
		\hline
		\hline
		\multicolumn{2}{c}{Date} & Eff. Time & Zd & Spectral & Obs.\\
		\cline{1-2}
		[yyyy mm dd] & [MJD] & [hr] & [$\deg$] & State & conditions\\
		\hline
		 2007 07 13 & 54294 & 1.78 & 6.5-17.0 & \\
		 2007 09 19 & 54362 & 0.71 & 25.1-50.8 &\\
		 2007 09 20 & 54363 & 1.43 & 21.3-40.9 &\\
		 2007 10 05 & 54378 & 0.85 & 6.5-26.4 &\\
		 2007 10 06 & 54379 & 1.85 & 6.4-25.8 &\\ 
		 2007 10 08 & 54381 & 1.95 & 17.8-43.1 & HS & MONO$_{\mathrm{wobble}}$\\
		 2007 10 09 & 54382 & 0.77 & 9.6-34.3 &\\
		 2007 10 10 & 54383 & 2.26 & 6.9-33.3 &\\
		 2007 10 11 & 54384 & 0.76 & 11.1-33.3 &\\
		 2007 11 05 & 54409 & 0.58 & 34.2-48.6 &\\
		 2007 11 06 & 54410 & 0.96 & 20.0-33.2 &\\
		 \hline
		 2008 07 02 & 54649 & 4.24 & 6.5-30.1 &\\
		 2008 07 03 & 54650 & 3.26 & 6.5-30.3 &\\
		 2008 07 04 & 54651 & 4.27 & 6.5-30.1 &\\
		 2008 07 05 & 54652 & 4.15 & 6.4-36.1 &\\
		 2008 07 06 & 54653 & 3.75 & 6.5.36.3 & HS & MONO$_{\mathrm{on/off}}$\\
		 2008 07 07 & 54654 & 3.69 & 6.5-37.4 &\\
		 2008 07 08 & 54655 & 3.94 & 6.5-34.1 &\\
		 2008 07 09 & 54656 & 3.06 & 6.5-33.8 &\\
		 2008 07 10 & 54657 & 2.89 & 6.5-36.8 &\\
		 2008 07 11 & 54658 & 1.18 & 6.5-30.1 &\\
		 \hline
		 2008 07 09 & 54656 & 0.33 & 28.5-33.5 &\\
	     2008 07 10 & 54657 & 0.39 & 21.5-36.5 &\\
	     2008 07 11 & 54658 & 0.32 & 14.8-19.6 &\\
         2008 07 12 & 54659 & 2.51 & 6.5-31.0 & \\
         2008 07 24 & 54671 & 0.62 & 13.0-19.6 &\\
         2008 07 25 & 54672 & 0.63 & 8.4-14.4 & HS & MONO$_{\mathrm{wobble}}$\\
         2008 07 26 & 54673 & 0.84 & 6.5-9.1 &\\
         2008 07 27 & 54674 & 0.30 & 9.5-12.7 &\\	
		 2009 06 30 & 55012 & 3.50 & 6.0-30.0 &\\
		 2009 07 01 & 55013 & 2.63 & 6.0-30.0 & \\
		 2009 07 02 & 55014 & 1.83 & 6.0-30.0 &\\
		 2009 07 05 & 55017 & 0.22 & 25.0-35.0 &  \\
		 \hline	
		 2009 10 08 & 55112 & 0.26 & 6.1-14.3 &\\
		 2009 10 10 & 55114 & 0.67 & 20.0-32.6 &\\
		 2009 10 11 & 55115 & 2.03 & 6.0-40.4 & \\
		 2009 10 12 & 55116 & 2.34 & 6.9-42.4 & \\
		 2009 10 13 & 55117 & 0.95 & 26.0-41.2 &\\
		 2009 10 14 & 55118 & 1.98 & 7.5-40.0 &\\
		 2009 10 16 & 55120 & 1.37 & 7.5-40.0 &\\
		 2009 10 17 & 55121 & 0.96 & 7.5-40.0 & \\
 		 2009 10 18 & 55122 & 1.60 & 7.5-40.0 & HS & STEREO$_{\mathrm{pre}}$\\
		 2009 10 19 & 55123 & 0.68 & 7.5-40.0 & \\
		 2009 10 21 & 55125 & 1.99 & 7.5-40.0 &\\
		 2009 11 06 & 55141 & 0.37 & 7.5-40.0 &\\		 
		 2009 11 07 & 55142 & 0.64 & 7.5-40.0 &\\
 		 2009 11 13 & 55148 & 0.89 & 7.5-40.0 &\\
		 2010 03 26 & 55281 & 0.78 & 38.5-50.0 &\\ 
		 2011 05 12 & 55693 & 1.35 & 12.3-42.1 & \\
		 2011 05 13 & 55694 & 1.20 & 9.1-29.0 &\\
		 \hline
		 2014 09 17 & 56917 & 2.55 & 6.8-38.4 &\\
		 2014 09 18 & 56918 & 1.29 & 6.3-26.5 &\\
		 2014 09 20 & 56920 & 2.38 & 6.0-38.0 & SS & STEREO$_{\mathrm{post}}$\\
		 2014 09 23 & 56923 & 3.00 & 6.0-39.0 &\\
		 2014 09 24 & 56924 & 3.26 & 6.6-37.5 &\\
		 2014 09 25 & 56925 & 1.81 & 6.2-39.0 &\\
		 \hline
	\end{tabular}

\end{center}
\end{table*}

\section{Observations and Data Analysis}
\label{section2}
MAGIC is a stereoscopic system consisting of two 17 m diameter imaging atmospheric Cherenkov telescopes (IACTs) located in El Roque de los Muchachos in the Canary island of La Palma, Spain ($28.8\deg$N, $17.8\deg$ W, 2225 m a.s.l.). Until 2009, MAGIC consisted of just one stand-alone IACT  with an integral flux sensitivity about 1.6\% of the Crab Nebula flux in 50 hr of observation \citep{2009APh....30..293A}. After autumn 2009, the second telescope (MAGIC\,II) started operation, allowing us to reach an energy threshold as low as 50 GeV at low zenith angles \citep{2012APh....35..435A}. In this period the sensitivity improved to $0.76\pm0.03$\% of the Crab Nebula flux for energies greater than 290 GeV in 50 hr of observations. Between summer 2011 and 2012 both telescopes underwent a major upgrade that involved the digital trigger, readout systems and the MAGIC\,I camera \citep{2016APh....72...61A}. After this upgrade, the system achieves, in stereoscopic observational mode, an integral sensitivity of $0.66\pm0.03$\% of the Crab Nebula flux in 50 hr above 220 GeV \citep{2016APh....72...76A}.\\
The data analysis presented in this paper was carried out using the standard MAGIC analysis software (MARS; \citealt{Zanin2013}). Integral and differential flux upper limits (ULs) were computed making use of the full likelihood analysis developed by \cite{2012JCAP...10..032A}, which takes into account the different instrument response functions (IRFs) of the telescopes along the years, assuming a 30\% systematic uncertainty.\\

At La Palma, Cyg\,X-1 culminates at a zenith angle of $6\deg$. Observations, performed up to $50\deg$, were carried out in a stand-alone mode (with just MAGIC\,I) from July 2007 to summer 2009, and, in stereoscopic mode, from October 2009 up to October 2014. Two data taking modes were used: the false-source tracking mode called \textit{wobble-mode} and the \textit{on-off mode}. In the former one, MAGIC points at two or four different positions situated $0.4\deg$ away from the source to evaluate the background simultaneously \citep{Fomin1994}. In the latter mode, the \textit{on region} (where the signal from the source is expected) and the \textit{off region} (background signal) are observed separately. In this case, the background sample is recorded under same conditions (same epoch, zenith angle and atmospheric conditions) as for the \textit{on data} but with no candidate source in the field of view.
The total Cyg\,X-1 data sample recorded by MAGIC amounts to $\sim97$ hr after data quality cuts (62.5 hours in stand-alone mode, 20.1 hours during pre-upgrade stereo period and 14.3 hours post-upgrade). The data set was distributed over 53 nights between July 2007 and October 2014. The whole data sample extends over five yearly campaigns, characterized by different performances of the telescopes. Because of this, each epoch was analyzed separately with appropriate MC-simulated gamma-ray events. The details of the observations for each campaign are summarized in Table~\ref{table:1}. For convenience, the following code is used in the table to describe the different observational features: STEREO stands for stereoscopic mode while MONO is used when only MAGIC I was operating. In the latter, the subscript specifies the observational mode: \textit{on-off} or \textit{wobble mode}. In STEREO, only \textit{wobble mode} was used, so the subscript is used to specify whether the observations were taken before (pre) or after (post) the MAGIC upgrade. \\

Different criteria to trigger observations were used during the campaign to optimize observations, aimed at observing the system in a state, HS, similar to that in which we previously reported a possible detection. The X-ray spectral states were defined by using public \textit{Swift}-BAT (15-50 keV) and \textit{RXTE}-ASM (1.5-12 keV) data, except for the data taken in 2014 where only \textit{Swift}-BAT was considered (since \textit{RXTE}-ASM ceased science operations on January 3 2012). Between July and November 2007, the criteria used to prompt the observation was a \textit{Swift}-BAT flux larger than 0.2 counts cm$^{-2}$ s$^{-1}$ and a ratio between \textit{RXTE}-ASM one-day average (in counts s$^{-1}$ in a \textit{Shadow Scanning Camera} (SSC)) and \textit{Swift}-BAT lower than 200. This criterion is in agreement with the one set by \cite{Grinberg2013} to define the X-ray states of Cyg\,X-1 using  \textit{Swift}-BAT data: above 0.09 counts cm$^{-2}$ s$^{-1}$  the microquasar stays in the HS+IS and below in the SS. The trigger criterion we selected  is higher to achieve a count rate similar to that of the previous \textit{MAGIC hint}. In July 2008, on top of the HS triggering criteria described above, we intensified observations following the X-ray superorbital modulation. The observations were triggered when the source was on the same superorbital phase as during the hint. Between June and October 2009, a new hardness ratio constraint using \textit{RXTE}-ASM data of the energy ranges 5-12 keV and 1.3-2 keV was included: the observations were only stopped after 5 consecutive days of this ratio being lower than 1.2, to avoid interrupting the observations during the IS. In May 2011, the source was observed on two nights based on internal analysis of public \textit{Fermi}-LAT data that showed a hint at HE during a hard X-ray activity period. Since all the above mentioned data were taken during the HS, for completeness, Cyg\,X-1 was also observed in its SS in September 2014 to exclude gamma-ray emission in this state at the same flux level as in the previous one. To define the X-ray state of the source, \textit{Swift}-BAT public data was again used following \cite{Grinberg2013} criteria.

\section{Results}
\label{section3}
\subsection{Search for steady emission}
We searched for steady VHE gamma-ray emission from Cyg\,X-1 at energies greater than 200 GeV making use of the entire data set of almost 100 hr. No significant excess was achieved. We computed ULs assuming a simple power-law function with different photon indices. The lower value, $\Gamma=2$, is consistent with the results obtained in the HE band (\citealt{Zanin2016}, \citealt{2016arXiv160705059Z}), while the upper one, $\Gamma=3.8$, is constrained by the former MAGIC results ($\Gamma=3.2\pm0.6$, \citealt{Albert2007}). Deviations in the photon index do not critically affect our results, quoted in Table~\ref{table:integralULsGamma}. Therefore all ULs obtained in this work are given at a confidence level (CL) of 95\% with $\Gamma=3.2$, which is the photon index obtained for the \textit{MAGIC hint}. For steady emission, we obtain an integral flux UL for energies greater than 200 GeV of $2.6\times10^{-12}$~photons cm$^{-2}$s$^{-1}$. Differential flux ULs for the entire data sample can be found in Table~\ref{table:differentialULs}. 

\subsubsection{Results during Hard State}
Observations under this X-ray spectral state were carried out between July 2007 and May 2011 reaching $\sim 83$ hr, where different criteria for triggering observations were used (see Section \ref{section2}). 
No significant excess was detected during this spectral state. The integral flux UL for energies greater than 200 GeV is $2.6\times10^{-12}$~photons cm$^{-2}$s$^{-1}$. Differential flux ULs are listed in the upper part of Table~\ref{table:differentialULs_states}. 
In order to search for VHE orbital modulation, we carried out an orbital phase-folded analysis. To accomplish a good compromise between orbital phase resolution and significant statistics, the binning in this analysis was 0.2. Moreover, in order to cover the superior conjunction of the BH (phases 0.9--0.1), we started to bin the data at phase 0.1. No VHE orbital modulation is evident either. Integral UL for a phase-folded analysis are shown in Table~\ref{table:phase-wise}. 

\begin{table}
\begin{center}
\caption{UL to the integral flux above 200 GeV at 95\% CL assuming a power-law spectrum with different photon indices, $\Gamma$.} 
\label{table:integralULsGamma}      
	\begin{tabular}{c c}
		\hline
		\hline
		$\Gamma$ & Flux UL at 95\% CL\\
		\cline{2-2}
	    & [$\times10^{-12}$~photons cm$^{-2}$s$^{-1}$]\\
		\hline
		2.0 & 2.20\\
    		2.6 & 2.44\\	
    		3.2 & 2.62\\	
   		3.8 & 2.71\\	
		\hline
	\end{tabular}
	\end{center}
\end{table}

\subsubsection{Results during Soft State}
Cyg\,X-1 was observed for a total of $\sim 14$ hr in the SS, bringing forth a clear difference on effective time with respect to the HS. Nevertheless, this corresponds to the post-upgrade period, in which MAGIC achieved its best sensitivity, $0.66\pm0.03$\% of the Crab Nebula flux above 220 GeV in 50 hr \citep{2016APh....72...76A}, implying that the flux sensitivity of previous observations was nearly reached in only about 9 hr. This data set guarantees, in turn, a full coverage of the X-ray spectral states that the source usually exhibits. Although steady gamma-ray emission in the SS, when no persistent jets are present, is not theoretically predicted, transient jet emission cannot be dismissed during this state, as it happens in the case of Cygnus\,X-3 (\citealt{2009Natur.462..620T}, \citealt{2009Sci...326.1512F}). Nevertheless, we did not find significant VHE gamma-ray emission from Cyg\,X-1 in the SS. Integral UL for energies beyond 200 GeV and $\Gamma=3.2$ is $1.0\times10^{-11}$~photons cm$^{-2}$s$^{-1}$. Differential ULs are quoted in the lower part of Table~\ref{table:differentialULs_states}. The integral ULs for the orbital phase-folded study are also given in Table~\ref{table:phase-wise}.

\subsection{Search for variable emission}
\begin{figure*}
		\centering
		\includegraphics[width=\textwidth,height=20cm]{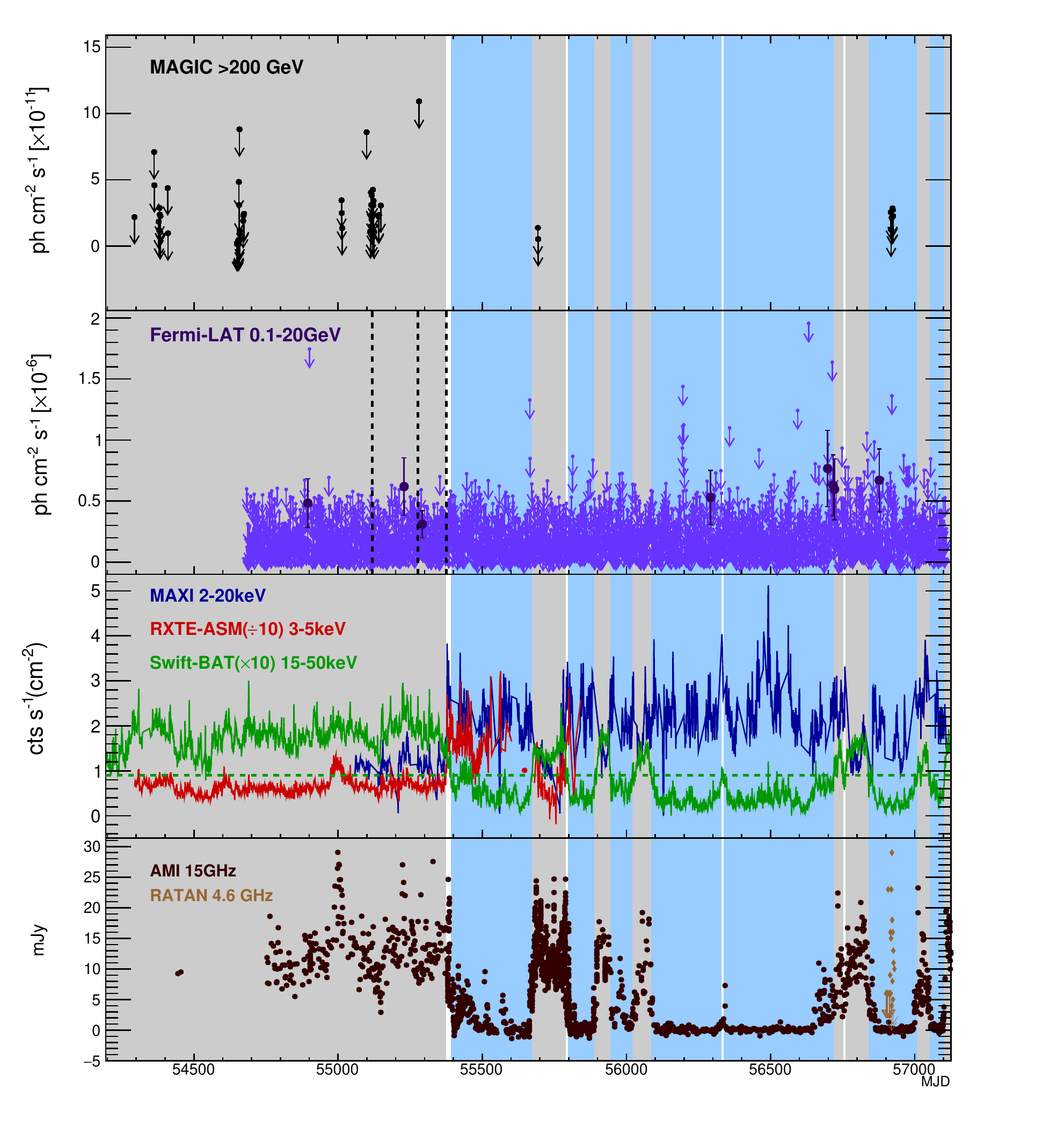}
		\caption{\textit{From top to bottom:} daily MAGIC integral ULs for $E > 200$ GeV assuming a power-law function with photon index $\Gamma=3.2$, HE gamma rays from \textit{Fermi}/LAT given by \protect\cite{Zanin2016}, hard X-ray (\textit{Swift}/BAT, $\times10$ counts s$^{-1}$cm$^{-2}$ in the 15--50 keV range), intermediate-soft X-ray (\textit{MAXI}, in counts s$^{-1}$ in the 2--20 keV range), soft X-ray (\textit{RXTE}/ASM, counts s$^{-1}$ divided by 10 in the 3--5 keV range), and finally, radio integral fluxes from AMI at 15 GHz and RATAN-600 at 4.6 GHz. In the HE pad, daily fluxes with $TS > 9$ are displayed as filled black points while days with $TS < 9$ are given as 95\% CL ULs. Dashed lines, in the same pad, correspond to \textit{AGILE} alerts. For convenience, an horizontal green dashed line in \textit{Swift}/BAT plot is displayed at the limit of 0.09 counts cm$^{-2}$s$^{-1}$, above which the source can be considered to be in the HS and below which it is in the SS (\protect\citealt{Grinberg2013}). This distinction between X-ray states is also highlighted by the color bands: gray bands correspond to the HS+IS and blue ones to the SS periods. White bands correspond to transitions between these two main X-ray spectral states which cannot be included within the HS periods. Zoomed view of MAGIC periods around MJD 55012--55281, MJD 55693--55694 and MJD 56917--56925 are shown in Fig.~\ref{fig:MW_LC_zoom5}, \ref{fig:MW_LC_zoom6} and \ref{fig:MW_LC_zoom9}, respectively.  }
		\label{fig:MW_LC}
\end{figure*}

Taking into account the X-ray and radio variability detected in Cyg\,X-1, as well as the rapid variation of the flux level previously reported by MAGIC on a timescale of hours, we carried out daily analysis for the 53 nights. This search yielded no significant excess. Integral ULs (95\% CL) for energies above 200 GeV for single-night observations are listed in Table~\ref{table:IntegralULs}.\\
MAGIC results are included in the top panel of the multiwavelength light curve presented in Fig.~\ref{fig:MW_LC} (zoom views are depicted in Fig.~\ref{fig:MW_LC_zoom5}, \ref{fig:MW_LC_zoom6} and \ref{fig:MW_LC_zoom9}). Besides MAGIC ULs, the figure shows data in the HE gamma-ray regime from \textit{Fermi}/LAT (0.1-20 GeV) given by \citealt{Zanin2016}, hard X-ray (\textit{Swift}/BAT in 15-50 keV, \citealt{Krimm2013}), intermediate-soft X-ray (\textit{MAXI} between 2-20 keV, \citealt{Matsuoka2009}), soft X-ray (quick-look results provided by the \textit{RXTE}/ASM team in 3-5 keV) and radio data (AMI at 15 GHz and RATAN-600 at 4.6 GHz). The three transient episodes observed by \textit{AGILE} are also marked. 

During this multi-year campaign, Cyg\,X-1 did not display any X-ray flare like that in which the \textit{MAGIC hint} was obtained. This prevented us from observing the source under strictly the same conditions: the maximum \textit{Swift}-BAT flux simultaneous to our observations happened on MJD 54379 ($1.13\sigma$, around superior conjunction of the BH) at the level of 0.23 counts cm$^{-2}$ s$^{-1}$, close but still lower than 0.31 counts cm$^{-2}$ s$^{-1}$ peak around the \textit{MAGIC hint}. However, we observed Cyg\,X-1 in coincidence with the first \textit{AGILE} flare. This transient episode (on October 16 2009, MJD 55120) showed $\sim4.1\sigma$ between 0.1-3 GeV with a gamma-ray flux of $(2.32\pm0.66)\times10^{-6}$~photons cm$^{-2}$s$^{-1}$ \citep{2010ApJ...712L..10S}, which took place during the X-ray HS of Cyg\,X-1. The corresponding MAGIC integral flux UL above 200 GeV for this day is $1.3\times10^{-11}$~photons cm$^{-2}$s$^{-1}$ (see Table~\ref{table:IntegralULs}). It is worth noting that \textit{Fermi}-LAT did not detect any significant signal in the energy range of 0.1-20 GeV on or around this date either \citep{Zanin2016}. The apparent discrepancy of \textit{Fermi}-LAT and \textit{AGILE} could be explained based on the different exposure time and off-axis angle distance both satellites presented during Cyg\,X-1 observations, as explained by \cite{2016ApJ...829..101M}. 

\begin{center}
\begin{table}
\caption{Differential flux ULs at 95\% CL for the overall data sample assuming a power-law spectrum with photon index of $\Gamma=3.2$.} 
\hfill{}
\label{table:differentialULs}      
	\begin{tabular}{ c c c}
		\hline
		\hline
		Energy range & Significance & Differential flux UL for $\Gamma=3.2$\\
		 \hline
		[GeV] & [$\sigma$] & [$\times10^{-13}$ TeV$^{-1}$cm$^{-2}$s$^{-1}$]\\
		\hline
		186--332 & 2.15 & 0.02\\
		332--589 & $-$0.14 & 0.33\\
		589--1048 & 0.44 & 0.18\\
		1048--1864 & 0.17 & 6.41\\
		1864--3315 & 0.03 & 75.64\\
		\hline
	\end{tabular}
	\hfill{}
\end{table}
\end{center}

\begin{figure*}
	\includegraphics[width=0.75\textwidth]{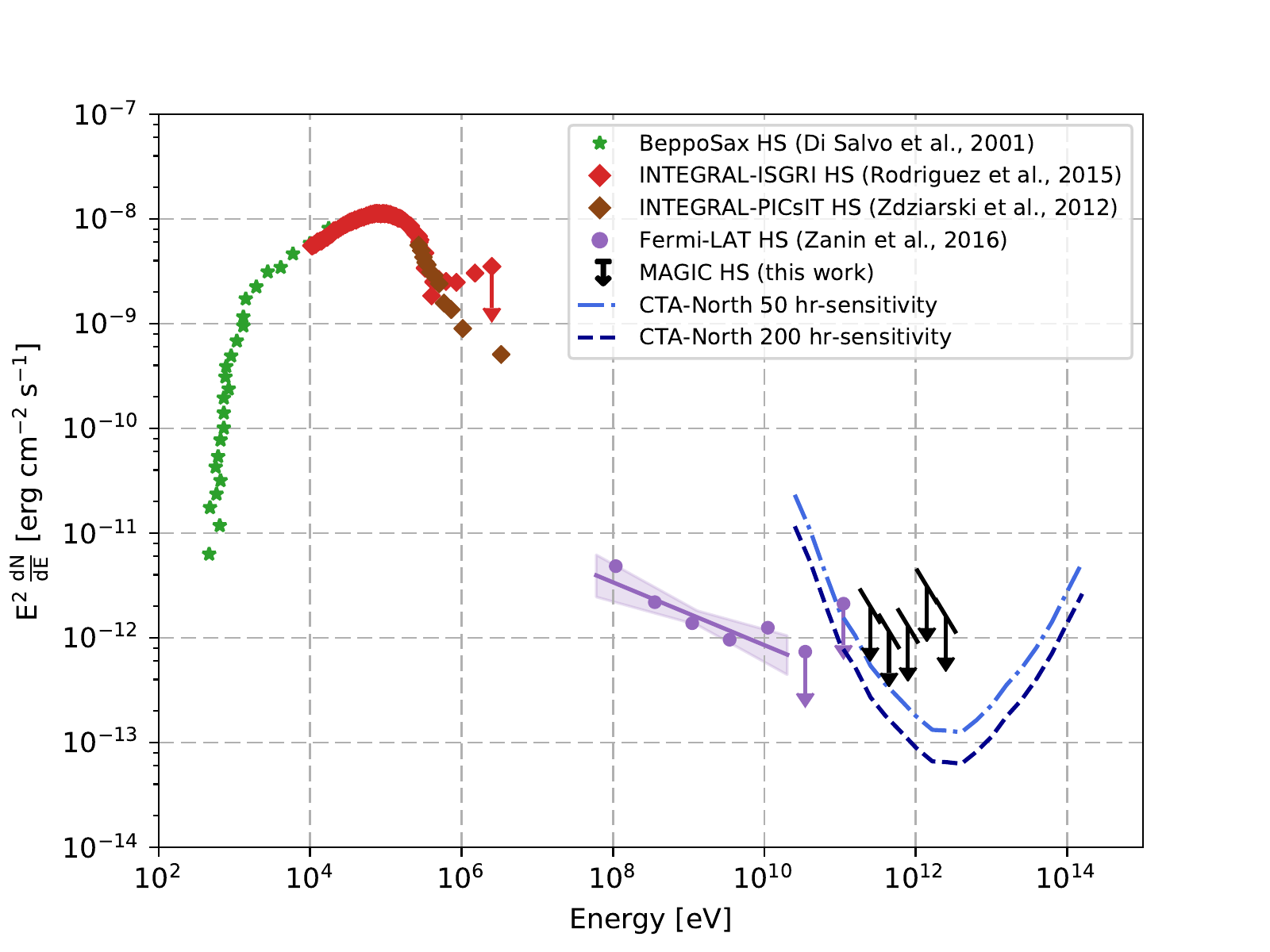}
\centering	
		\caption{Spectral energy distribution (SED) of Cyg\,X-1 covering X-ray, HE and VHE gamma-ray regimes during the HS. \textit{BeppoSAX} soft X-ray data (in the keV band, green stars) is taken from \protect\cite{2001ApJ...547.1024D}, while for the hard X-ray band, data from both \textit{INTEGRAL}-ISGRI (10 keV-2 MeV, red diamond and UL; \protect\citealt{2015ApJ...807...17R}) and \textit{INTEGRAL}-PICsIT  (150 keV-10 MeV, brown diamond; \protect\citealt{Zdziarski2012}) are displayed, given their incompatibility spectral results above 1 MeV.  In the HE gamma-ray band (60 MeV-few hundred GeV, violet circles and ULs), results from \protect\cite{Zanin2016} obtained with \textit{Fermi}-LAT data are shown, including its best fit (power law with photon index $\Gamma=2.3\pm0.1$). At VHE, results from this work during the HS are plotted (black) assuming a power-law function of $\Gamma=3.2$. The dashed blue lines correspond to the 50 and scaled to 200 hr sensitivity curves for CTA North. No statistical errors are drawn, except for the \textit{Fermi}-LAT butterfly.}
		\label{fig:SED}
\end{figure*}

\section{Discussion}
VHE gamma-ray emission from microquasars has been proposed in the literature from both leptonic (e.g. \citealt{1999MNRAS.302..253A}, \citealt{2006A&A...447..263B}) and hadronic processes (e.g. \citealt{2003A&A...410L...1R}). The most efficient radiative process is inverse Compton (IC), although hadronic emission may also occur in dense matter or HE radiation environments (see \citealt{2009IJMPD..18..347B}, and references therein). There are different possible source photon fields according to the distance of the production site to the compact object: close to the BH, IC of thermal photons (\citealt{2002A&A...388L..25G}, \citealt{2002A&A...393L..61R}), or synchrotron photons \citep[e.g.][]{2006A&A...447..263B} may be dominant. When the production region is situated inside the binary but further from the BH, the process can take place on photons from the companion star. In fact, anisotropic IC on stellar photons likely taking place in the jet seems to be the main mechanism of HE gamma-ray production in Cyg\,X-1 (\citealt{Zanin2016}; see \citealt{2016arXiv160705059Z} for additional possible contributions in gamma rays). Finally, in this source VHE gamma-ray emission may be also produced in the region where the jets seem to interact with the environment \citep{2005Natur.436..819G}, as proposed for instance by \cite{2009A&A...497..325B}. 

In the first two cases, i.e. if VHE emission is produced inside the binary system Cyg\,X-1, the VHE photons will suffer severe absorption through pair creation in the stellar photon field (e.g. \citealt{2007A&A...476....9O}, \citealt{2007A&A...464..437B}). This absorption is modulated due to changes in the star-emitter-observer relative positions along the orbit, with the maximum (minimum) of the attenuation, and the lowest (highest) energy threshold, taking place at the superior (inferior) conjunction of the compact object, which corresponds to phase 0 (0.5) in Cyg\,X-1. If orbitally modulated VHE emission were detected, it would likely imply that this emission comes at most from the outskirts of the binary system, approximately between $10^{12}$ and $10^{13}$ cm from the BH (see \citealt{2008A&A...489L..21B}), a location still consistent with the constraints derived from the GeV data \citep{Zanin2016}. As in the case of gamma-ray absorption through pair creation, geometric effects are also relevant for IC processes, with the maximum probability of interaction between electrons and stellar photons occurring at superior conjunction of the compact object and the minimum at inferior conjunction. Further out of this region ($> 10^{13}$ cm), VHE emission would be less affected by orbital motion, although particle acceleration and IC cooling are expected to be also weaker there, which may mean little or no production of VHE photons.\\

MAGIC observations carried out between July 2007 and September 2014 for a total of $\sim100$ hours covered the two principal X-ray states of Cyg\,X-1 with the main focus on the HS, where the source has shown to accelerate relativistic particles that produce GeV gamma rays likely coming from the jets \citep{Zanin2016}. We did not detect any significant excess from either all the data or any of the samples, including orbital phase-folded and daily analysis. In this long-term campaign, we provided, for the first time, constraining ULs on the VHE emission of Cyg\,X-1 at the two main X-ray states, the HS and the SS, separately as well as in an orbital binning base, which showed no hint of gamma-ray orbital modulation. This was possible thanks to a comprehensive trigger strategy that allowed us to observe the source under flaring activity. The chosen photon index ($\Gamma=3.2$ in this work, Crab-like in the previous MAGIC observations, \citealt{Albert2007}) and the addition of 30\% systematic uncertainties contributed to obtain more robust ULs compared to the formerly ones reported by MAGIC. 

The total power emitted by the jets during the HS in Cyg\,X-1 is expected to be $10^{36}-10^{37}$ erg s$^{-1}$ \citep{2005Natur.436..819G}. The integral UL $2.6\times10^{-12}$~photons cm$^{-2}$s$^{-1}$, for energies greater than 200 GeV, obtained by MAGIC in this work corresponds to a luminosity of $6.4\times10^{32}$ erg s$^{-1}$ assuming a distance of 1.86 kpc \citep{Reid2011}. Therefore, the UL on the conversion efficiency of jet power to VHE gamma ray luminosity is 0.006--0.06\%, similar to the one obtained for Cyg\,X-3 \citep{2010ApJ...721..843A}. Note that gamma-ray opacity in Cyg\,X-3 is nevertheless about two orders of magnitude higher than in Cyg\,X-1. 

VHE emission from the jet large scale or jet-medium interaction regions above the sensitivity level of MAGIC can be ruled out, as these regions are not affected by gamma-ray absorption. On the binary scales, however, the non-detection is less conclusive because of pair creation in the stellar photon field. Models do predict VHE radiation as long as particle acceleration is efficient (e.g. \citealt{Pepe2015}). Formally, particle acceleration up to $\sim$~TeV energies can be reached in the jet on the binary region \citep{2008MNRAS.383..467K}, and thus 100~GeV IC photons should be produced, but this emission may be right below the detection level of MAGIC (as in \citealt{2016arXiv160705059Z}, Fig. 6) even under negligible gamma-ray absorption. It could otherwise be that non-thermal particles cannot reach VHE IC emitting energies in the jet of Cyg\,X-1. Besides inefficient acceleration, a very high magnetic field could also prevent particles to reach VHE, and even if these particles were present, a strong magnetic field can suppress intensely VHE photon production.

Nevertheless, one cannot dismiss the possibility of a transient emission as the one hinted by MAGIC in 2006. This flare took place during an orbital phase around the superior conjunction of the BH, where the gamma-ray absorption is expected to be the highest. The attenuation constraint may have been relaxed by an emitter at some distance from the BH \citep{Albert2007}, with its intrinsic variability possibly related for instance to jet-stellar wind interaction (\citealt{2008A&A...482..917P}, \citealt{2009ApJ...696..690O}). On the other hand, even considering absorption by stellar photons, emission closer to the BH would be possible accounting for extended pair cascades under a reasonable intrinsic gamma-ray luminosity, although rather low magnetic fields in the stellar wind would be required (\citealt{Zdziarski2009}; see also \citealt{2008A&A...489L..21B}). Cyg\,X-3, the other microquasar firmly established as a GeV emitter (\citealt{2009Natur.462..620T}, \citealt{2009Sci...326.1512F}), displays a very different behaviour from that of Cyg\,X-1. The HE gamma-ray emission from Cyg\,X-3 is transient, occurring sometimes during flaring activity of non-thermal radio emission from the jets \citep{2012MNRAS.421.2947C}. If VHE radiation in microquasars were related to discrete radio-emitting-blobs with high Lorentz factor ($\Gamma\geq2$), this may also happen in Cyg\,X-1 during hard-to-soft transitions. 

The multiwavelength emission from X-rays up to VHE gamma rays in Cyg\,X-1 is shown in Fig.~\ref{fig:SED}. The data used in this spectral energy distribution (SED) corresponds to the HS. The sensitivity curve for 50 and scaled to 200 hr of observations with the future Cherenkov Telescope Array, CTA\footnote{as shown in \href{https://www.cta-observatory.org/science/cta-performance/}{https://www.cta-observatory.org/science/cta-performance/}}, on the Northern hemisphere is showed along with the data. The spectral cutoff of the HE radiation from Cyg\,X-1 is still unknown, although if the gamma-ray emission in the HS reaches $\sim$~TeV energies, the next generation of IACTs may be able to detect the system for long enough exposure times. Thus, to detect steady VHE emission from the jets, future more sensitive instruments, as CTA, would be needed. This instrument could provide valuable information of the VHE gamma-ray production in Cyg\,X-1 (HE spectral cutoff, energetics, impact of gamma-ray absorption/IC cascades), as well as allow the study of possible short-term flux variability.

\section*{Acknowledgements}

We would like to thank
the Instituto de Astrof\'{\i}sica de Canarias
for the excellent working conditions
at the Observatorio del Roque de los Muchachos in La Palma.
The financial support of the German BMBF and MPG,
the Italian INFN and INAF,
the Swiss National Fund SNF,
the ERDF under the Spanish MINECO
(FPA2015-69818-P, FPA2012-36668, FPA2015-68278-P,
FPA2015-69210-C6-2-R, FPA2015-69210-C6-4-R,
FPA2015-69210-C6-6-R, AYA2013-47447-C3-1-P,
AYA2015-71042-P, ESP2015-71662-C2-2-P, CSD2009-00064),
and the Japanese JSPS and MEXT
is gratefully acknowledged.
This work was also supported
by the Spanish Centro de Excelencia ``Severo Ochoa''
SEV-2012-0234 and SEV-2015-0548,
and Unidad de Excelencia ``Mar\'{\i}a de Maeztu'' MDM-2014-0369,
by the Croatian Science Foundation (HrZZ) Project 09/176
and the University of Rijeka Project 13.12.1.3.02,
by the DFG Collaborative Research Centers SFB823/C4 and SFB876/C3,
and by the Polish MNiSzW grant 745/N-HESS-MAGIC/2010/0. The AMI arrays are supported by the United Kingdom STFC
and by the University of Cambridge.  R.Z.  acknowledges  the  Alexander  von  Humboldt  Foundation
for the financial support and the Max-Planck Institut fur Kernphysik as hosting
institution. This research has made use of MAXI data provided by RIKEN, JAXA and the MAXI team.


\begin{center}
\begin{table*}
\caption{Differential flux ULs at 95\% CL for each X-ray spectral state.} 
\hfill{}
\label{table:differentialULs_states}      
	\begin{tabular}{c c c c}
		\hline
		\hline
		Spectral State & Energy range & Significance & Differential flux UL for $\Gamma=3.2$\\
		 \hline
		& [GeV] & [$\sigma$] & [$\times10^{-12}$ TeV$^{-1}$cm$^{-2}$s$^{-1}$]\\
		\hline
		& 186--332 & $-$2.57 & 0.20\\
		& 332--589 & $-$0.03 & 3.70\\
		HS & 589--1048 & 2.09 & 1.31\\
		& 1048--1864 & 0.02 & 99.22\\
		& 1864--3315 & 0.51 & 16.34\\
		\hline
		& 186--332 & 1.14 & 0.49\\
		& 332--589 & 1.22 & 0.11\\
		SS & 589--1048 & 0.06 & 4.71\\
		& 1048--1864 & $-$1.23 & 51.62\\
		& 1864--3315 & $-$1.34 & 16.37\\
		\hline
	\end{tabular}
	\hfill{}
\end{table*}
\end{center}

\begin{center}
\begin{table*}
\caption{Phase-wise 95\% CL integral flux ULs for energies $>200$ GeV for the HS and the SS observations. The latter did not cover phases from 0.9 to 0.1, so no ULs are provided.} 
\hfill{}
\label{table:phase-wise}      
	\begin{tabular}{c c c c c}
		\hline
		\hline
		Spectral State & Phase range & Eff. Time & Significance & Integral flux UL for $\Gamma=3.2$\\
		 \hline
		& & [hr] & [$\sigma$] & [$\times10^{-12}$~photons cm$^{-2}$s$^{-1}$]\\
		\hline
		& 0.1--0.3 & 15.47& $-$0.77  & 7.89\\
		& 0.3--0.5 & 22.34 & 1.88 & 6.91\\
		HS & 0.5--0.7 & 14.08&  0.00 & 21.32\\
		& 0.7--0.9 & 14.81 & 0.99 & 6.92\\
		& 0.9--0.1 & 15.62 & $-$0.96 & 4.34\\
		\hline
		& 0.1--0.3 & 2.58 & 0.45 & 19.32\\
		& 0.3--0.5 & 4.35 & $-$1.23 & 7.96\\
		SS & 0.5--0.7 & 3.91 & 0.59 & 15.49\\
		& 0.7--0.9 & 3.64 & 0.23 & 18.23\\
		& 0.9--0.1 & -- & -- & --\\
		\hline
	\end{tabular}
	\hfill{}
\end{table*}
\end{center}

\begin{table*}
\begin{center}
\caption{\textit{From left to right:} Date of the beginning of the observations in calendar and in MJD, effective time after quality cuts, significance for an energy threshold of $\sim150$ GeV for \textit{mono} observations (only MAGIC I) and $\sim100$ GeV for \textit{stereoscopic} observations (separated by the horizontal line) and integral flux ULs at 95\% CL for energies above 200 GeV computed on a daily basis. MJD 54656, 54657 and 54658 were analyzed separately according to each observational mode (see Table~\ref{table:1}). Due to low statistics, neither the integral UL for MJD  55017 nor the significant for MJD 55116 were computed.} 
\centering
        
\label{table:IntegralULs}      
	\begin{tabular}{ c c c c c}
		\hline
		\hline
		\multicolumn{2}{c}{Date} & Eff. Time & Significance & Flux UL for $\Gamma$=3.2\\		
		\cline{1-2} 
		[yyyy mm dd] & [MJD] & [hr] & [$\sigma$] & [$\times10^{-11}$~photons cm$^{-2}$s$^{-1}$] \\
		\hline
		 2007 07 13 & 54294 & 1.78 & -0.67 & 2.19 \\
		 2007 09 19 & 54362 & 0.71 &  1.10&  7.10  \\
		 2007 09 20 & 54363 & 1.43 & 1.99 &  4.59\\
		 2007 10 05 & 54378 & 0.85  & -0.84 & 1.84\\
		 2007 10 06 & 54379 & 1.85  & 0.02 & 1.21\\
		 2007 10 08 & 54381 & 1.95  & 0.99 & 2.88\\
		 2007 10 09 & 54382 & 0.77  & -0.57 & 2.38\\
		 2007 10 10 & 54383 & 2.26  & -0.04 & 1.05\\
		 2007 10 11 & 54384 & 0.76  & 1.68 & 2.26\\
		 2007 11 05 & 54409 & 0.58  & 0.31 & 4.38\\
		 2007 11 06 & 54410 & 0.96 & -1.24 & 0.97\\
		 2008 07 02 & 54649 & 4.24 & 2.33 & 0.21 \\
		 2008 07 03 & 54650 & 3.26 & 1.53 & 0.15\\
		 2008 07 04 & 54651 & 4.27 & 2.36 & 0.23\\
		 2008 07 05 & 54652 & 4.15 & 2.97 & 0.22\\
		 2008 07 06 & 54653 & 3.75 & 1.75 & 0.39\\
		 2008 07 07 & 54654 & 3.69 & 2.74 & 0.24\\
		 2008 07 08 & 54655 & 3.94 & 2.01 & 0.18\\
		 2008 07 09 & 54656 & 3.06 & 1.66 & 0.49\\
		 2008 07 10 & 54657 & 2.89 & 1.75 & 0.38\\
		 2008 07 11 & 54658 & 1.18 & 0.32 & 0.93\\
		 2008 07 09 & 54656 & 0.33 & 0.06 & 4.84\\
	     2008 07 10 & 54657 & 0.39 & -1.22 & 3.11\\
	     2008 07 11 & 54658 & 0.32 & 1.83 & 8.81\\
         2008 07 12 & 54659 & 2.51 & 0.11 & 1.16\\
         2008 07 24 & 54671 & 0.62 & -1.45 & 1.90\\
         2008 07 25 & 54672 & 0.63 & -0.15 & 2.30\\
         2008 07 26 & 54673 & 0.84 & -1.33 & 2.40\\
         2008 07 27 & 54674 & 0.30 & 2.09 & 2.44\\
		 2009 06 30 & 55012 & 3.50  & 0.76 & 3.46\\
		 2009 07 01 & 55013 & 2.63  & 0.73 & 2.50\\
		 2009 07 02 & 55014 & 1.83  & 0.14 & 1.36\\
		 2009 07 05 & 55017 & 0.22  & 0.37 & --\\
		 \hline
		 2009 10 08 & 55112 & 0.26  & -1.85 & 1.11 \\
		 2009 10 10 & 55114 & 0.67  & 0.19 & 1.50 \\
		 2009 10 11 & 55115 & 2.03  & 0.32 & 3.10 \\
		 2009 10 12 & 55116 & 2.34  & -- & 2.19\\
		 2009 10 13 & 55117 & 0.95 & 1.53 & 3.87\\
		 2009 10 14 & 55118 & 1.98  & -0.30 & 2.44\\ 
		 2009 10 16 & 55120 & 1.37 & -2.99 & 1.30 \\
		 2009 10 17 & 55121 & 0.96 & -0.77 & 4.25 \\
 		 2009 10 18 & 55122 & 1.60 & -0.27 & 3.05\\
		 2009 10 19 & 55123 & 0.68 & -0.44 & 3.42\\
		 2009 10 21 & 55125 & 1.99 & -1.90 & 1.09\\
		 2009 11 06 & 55141 & 0.37 & -3.04 & 2.23\\
		 2009 11 07 & 55142 & 0.64 & 0.13 & 2..35 \\
 		 2009 11 13 & 55148 & 0.89 & -1.23 & 3.06\\
 		 2010 03 26 & 55281 & 0.78 & 1.75 & 10.92\\		
		 2011 05 12 & 55693 & 1.35  & 0.09 & 1.38\\
		 2011 05 13 & 55694 & 1.20  & -1.54 & 0.53 \\
		 2014 09 17 & 56917 & 2.55 & 0.32 & 2.56 \\
		 2014 09 18 & 56918 & 1.29 & -0.99 & 1.25\\
		 2014 09 20 & 56920 & 2.38  & 0.08 &  2.13\\
		 2014 09 23 & 56923 & 3.00  & 0.85 & 2.85\\
		 2014 09 24 & 56924 & 3.26  & -0.61 & 2.73\\
		 2014 09 25 & 56925 & 1.81  & 0.28 & 2.26\\
		 \hline
	\end{tabular}

	\end{center}
\end{table*}

\clearpage


\begin{figure*}
		\centering
		\includegraphics[width=0.7\textwidth,height=11cm]{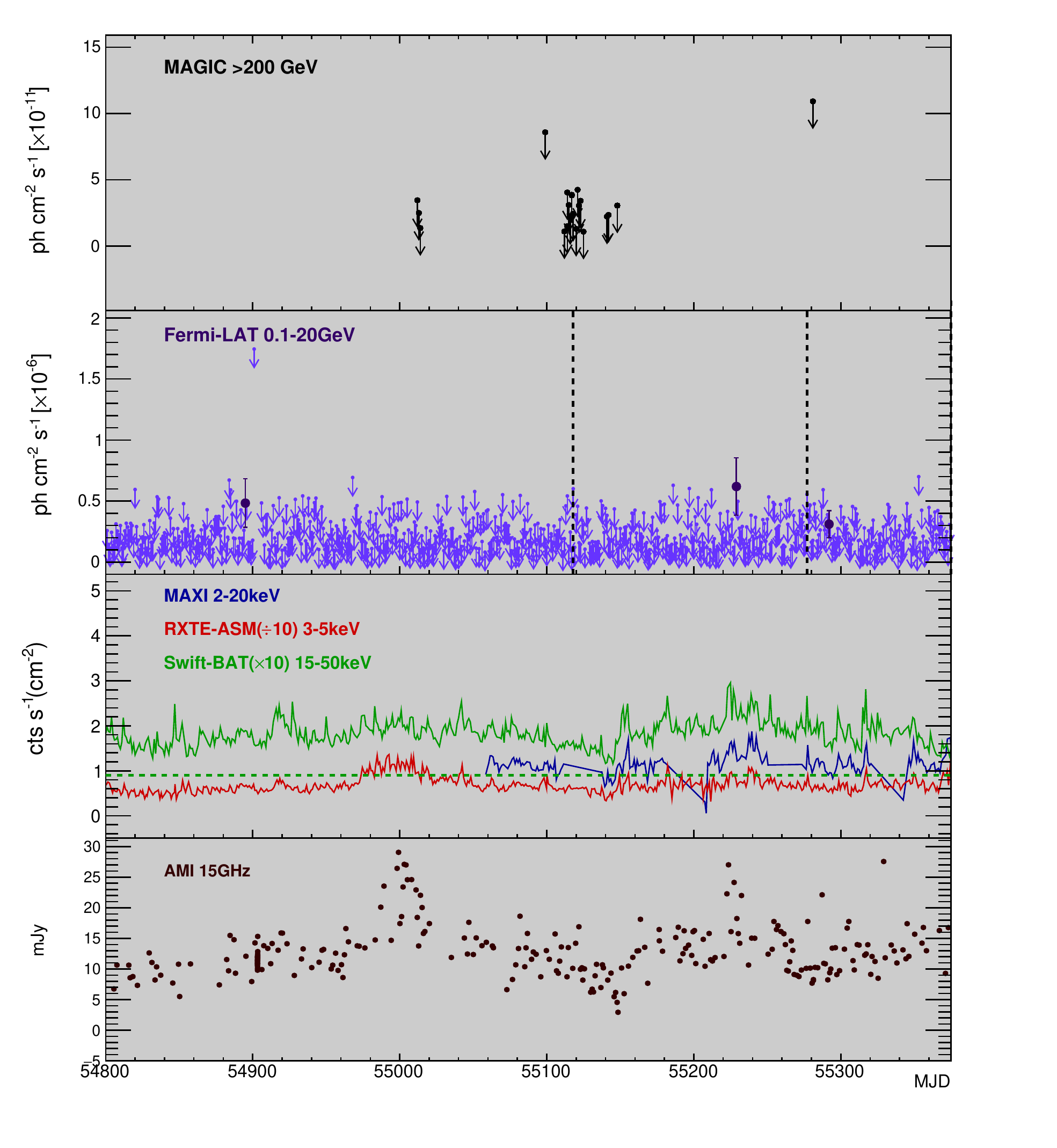}
		\caption{Zoomed view of Fig.~\ref{fig:MW_LC} around June 30 2009 (MJD 55012) to March 26 2010 (MJD 55281), corresponding to the HS of Cyg\,X-1.}
		\label{fig:MW_LC_zoom5}
\end{figure*}

\begin{figure*}
		\centering
		\includegraphics[width=0.7\textwidth,height=11cm]{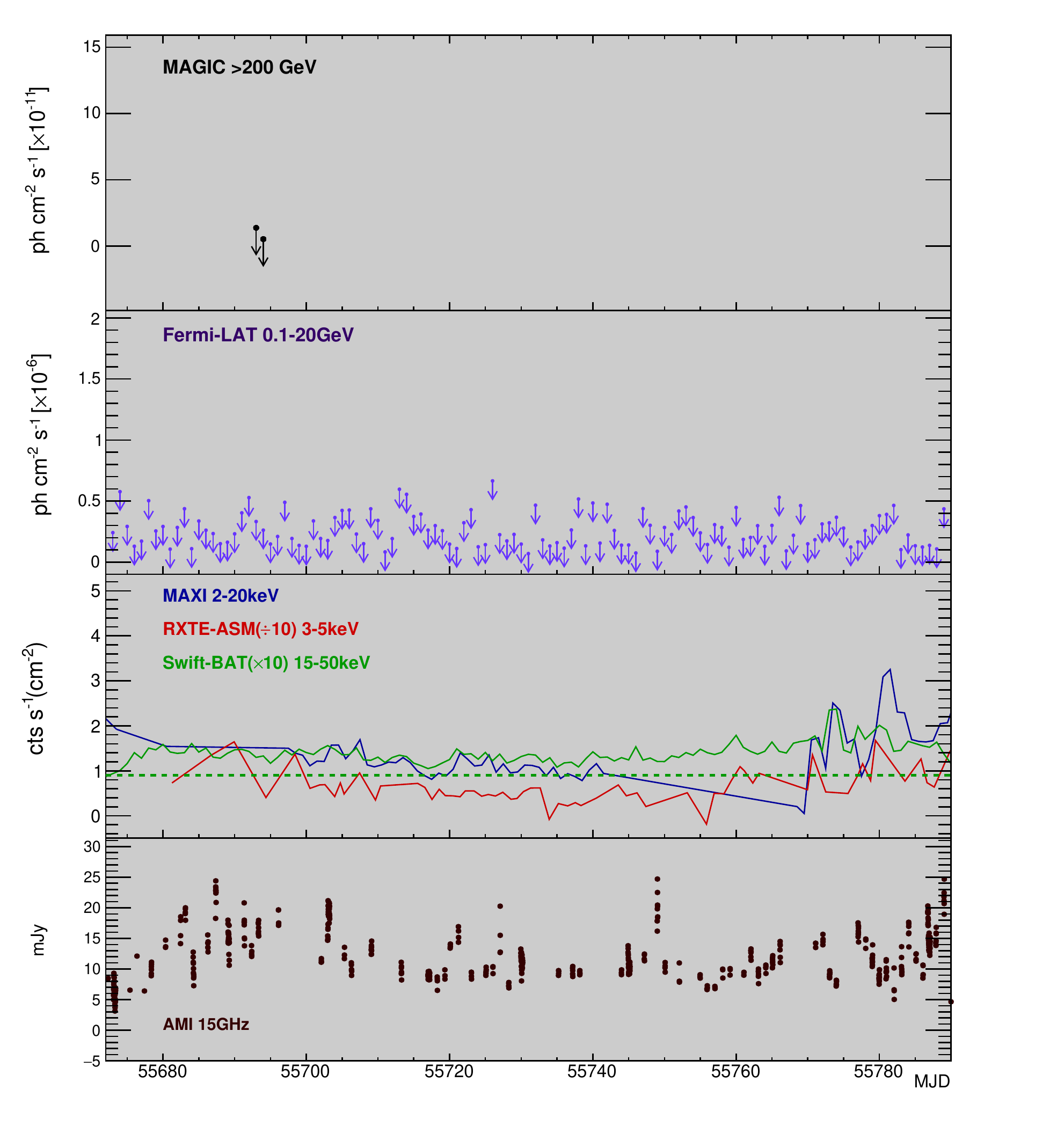}
		\caption{Zoomed view of Fig.~\ref{fig:MW_LC} around May 12 and 13 2011 (MJD 55693 and 55694, respectively), corresponding to the HS of Cyg\,X-1.}
		\label{fig:MW_LC_zoom6}
\end{figure*}

\begin{figure*}
		\centering
		\includegraphics[width=0.7\textwidth,height=11cm]{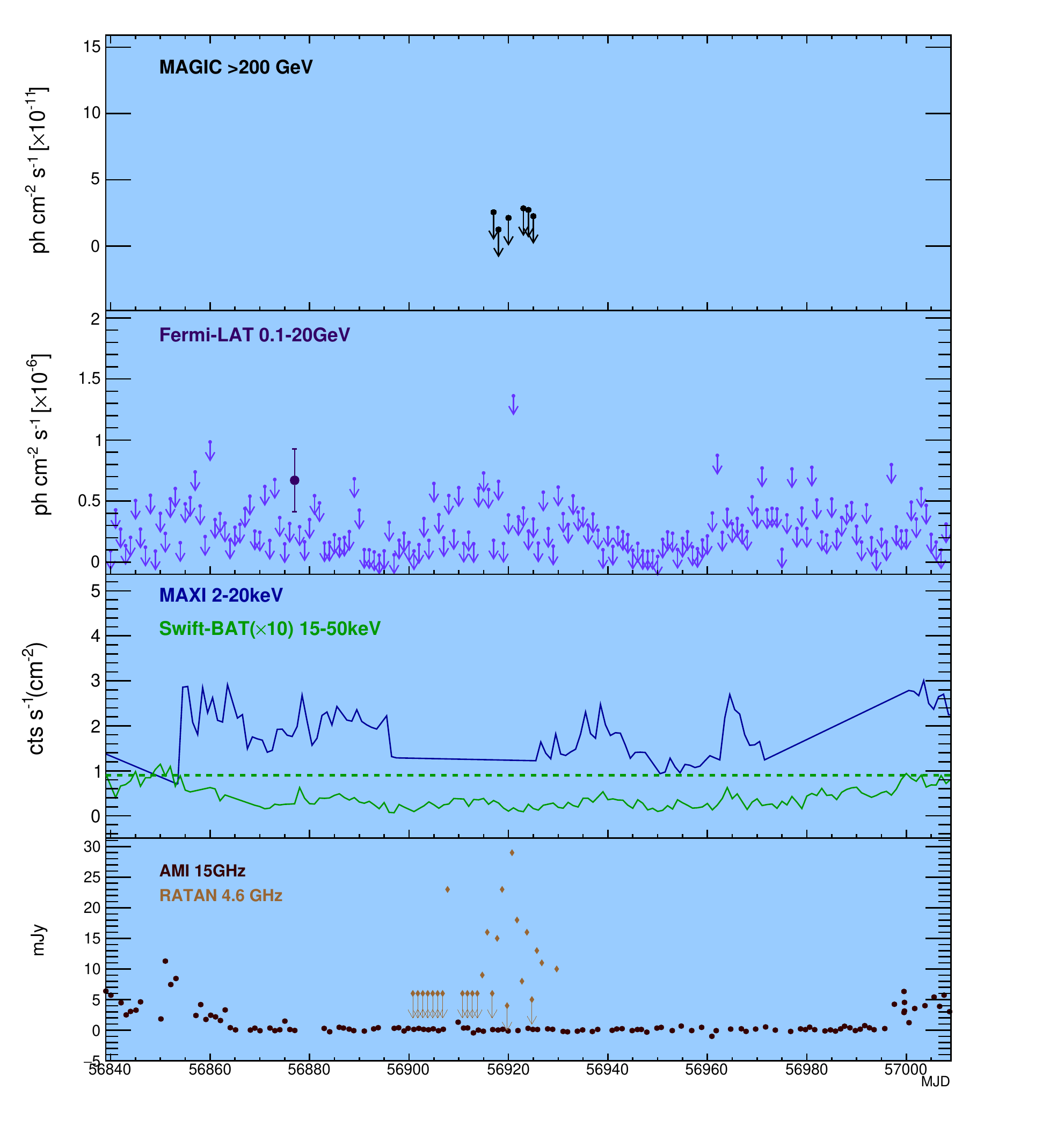}
		\caption{Zoomed view of Fig.~\ref{fig:MW_LC} around September 17 2014 (MJD 56917) to September 25 2014 (MJD 56925), corresponding to the SS of Cyg\,X-1.}
		\label{fig:MW_LC_zoom9}
\end{figure*}


\bibliographystyle{mnras}
\bibliography{bibliography_CygX1} 


\vspace*{0.5cm}
\noindent
$^{1}$ {ETH Zurich, CH-8093 Zurich, Switzerland} \\
$^{2}$ {Universit\`a di Udine, and INFN Trieste, I-33100 Udine, Italy} \\
$^{3}$ {INAF - National Institute for Astrophysics, viale del Parco Mellini, 84, I-00136 Rome, Italy} \\
$^{4}$ {Universit\`a di Padova and INFN, I-35131 Padova, Italy} \\
$^{5}$ {Croatian MAGIC Consortium, Rudjer Boskovic Institute, University of Rijeka, University of Split - FESB, University of Zagreb - FER, University of Osijek,Croatia} \\
$^{6}$ {Saha Institute of Nuclear Physics, 1/AF Bidhannagar, Salt Lake, Sector-1, Kolkata 700064, India} \\
$^{7}$ {Max-Planck-Institut f\"ur Physik, D-80805 M\"unchen, Germany} \\
$^{8}$ {Universidad Complutense, E-28040 Madrid, Spain} \\
$^{9}$ {Inst. de Astrof\'isica de Canarias, E-38200 La Laguna, Tenerife, Spain} \\
$^{10}$ {Universidad de La Laguna, Dpto. Astrof\'isica, E-38206 La Laguna, Tenerife, Spain} \\
$^{11}$ {University of \L\'od\'z, PL-90236 Lodz, Poland} \\
$^{12}$ {Deutsches Elektronen-Synchrotron (DESY), D-15738 Zeuthen, Germany} \\
$^{13}$ {Institut de Fisica d'Altes Energies (IFAE), The Barcelona Institute of Science and Technology, Campus UAB, 08193 Bellaterra (Barcelona), Spain} \\
$^{14}$ {Universit\`a  di Siena, and INFN Pisa, I-53100 Siena, Italy} \\
$^{15}$ {Institute for Space Sciences (CSIC/IEEC), E-08193 Barcelona, Spain} \\
$^{16}$ {Technische Universit\"at Dortmund, D-44221 Dortmund, Germany} \\
$^{17}$ {Universit\"at W\"urzburg, D-97074 W\"urzburg, Germany} \\
$^{18}$ {Finnish MAGIC Consortium, Tuorla Observatory, University of Turku and Astronomy Division, University of Oulu, Finland} \\
$^{19}$ {Unitat de F\'isica de les Radiacions, Departament de F\'isica, and CERES-IEEC, Universitat Aut\`onoma de Barcelona, E-08193 Bellaterra, Spain} \\
$^{20}$ {Universitat de Barcelona, ICC, IEEC-UB, E-08028 Barcelona, Spain} \\
$^{21}$ {Japanese MAGIC Consortium, ICRR, The University of Tokyo, Department of Physics, Kyoto University, Tokai University, The University of Tokushima, Japan} \\
$^{22}$ {Inst. for Nucl. Research and Nucl. Energy, BG-1784 Sofia, Bulgaria} \\
$^{23}$ {Universit\`a di Pisa, and INFN Pisa, I-56126 Pisa, Italy} \\
$^{24}$ {ICREA and Institute for Space Sciences (CSIC/IEEC), E-08193 Barcelona, Spain} \\
$^{25}$ {also at the Department of Physics of Kyoto University, Japan} \\
$^{26}$ {now at Centro Brasileiro de Pesquisas F\'isicas (CBPF/MCTI), R. Dr. Xavier Sigaud, 150 - Urca, Rio de Janeiro - RJ, 22290-180, Brazil} \\
$^{27}$ {now at NASA Goddard Space Flight Center, Greenbelt, MD 20771, USA} \\
$^{28}$ {Department of Physics and Department of Astronomy, University of Maryland, College Park, MD 20742, USA} \\
$^{29}$ {Humboldt University of Berlin, Institut f\"ur Physik Newtonstr. 15, 12489 Berlin Germany} \\
$^{30}$ {also at University of Trieste} \\
$^{31}$ {Japanese MAGIC Consortium} \\
$^{32}$ {now at Finnish Centre for Astronomy with ESO (FINCA), Turku, Finland} \\
$^{33}$ {also at INAF-Trieste and Dept. of Physics \& Astronomy, University of Bologna} \\
$^{34}$ {Departament d'Astronomia i Metereologia, Institut de Ci\`ences del Cosmos, Universtitat de Barcelona, Barcelona, Spain} \\
$^{35}$ {Cavendish Laboratory, J. J. Thomson Avenue, Cambridge CB3 0HE, UK} \\
$^{36}$ {Special astrophysical Observatory RAS, Nizhnij Arkhys, Karachaevo-Cherkassia, Russia } \\
$^{37}$ {Kazan Federal University, Kazan, Republic of Tartarstan, Russia} \\
$^{38}$ {Max-Planck-Institut fur Kernphysik, P.O. Box 103980, D 69029 Heidelberg, Germany} \\

\bsp	
\label{lastpage}
\end{document}